\documentclass[aps,pre,showpacs,twocolumn,superscriptaddress]{revtex4}
\usepackage{graphicx}
\usepackage{amssymb}
\usepackage{bm}% bold math

\begin{document}

\title{Robustness of retrieval properties against imbalance between long-term potentiation and depression of spike-timing-dependent plasticity}

\author{Narihisa Matsumoto}
\email[]{xmatumo@brain.riken.go.jp}
\affiliation{Graduate School of Science and Engineering, Saitama University, Saitama 338-8570, Japan}
\affiliation{RIKEN Brain Science Institute, Saitama 351-0198, Japan}

\author{Masato Okada}
\email[]{okada@brain.riken.go.jp}
\affiliation{RIKEN Brain Science Institute, Saitama 351-0198, Japan}
\affiliation{``Intelligent Cooperation and Control'', PRESTO, JST, Saitama 351-0198, Japan}

\date{\today}

\begin{abstract}
Spike-timing-dependent plasticity (STDP) has recently been shown
in some physiological studies.
STDP depends on the precise temporal relationship of pre- and
post-synaptic spikes.
Many authors have indicated that a precise balance between
long-term potentiation (LTP) and long-term depression (LTD) of
STDP is significant for a stable learning.
However, a situation in which the balance is maintained precisely
is inconceivable in the brain.
Using a method of the statistical neurodynamics, we show robust
retrieval properties of spatio-temporal patterns in an associative
memory model against the imbalance between LTP and LTD.
When the fluctuation of LTD is assumed to obey a Gaussian
distribution with mean $0$ and variance $\delta^2$, the storage
capacity takes a finite value even at large $\delta$.
This means that the balance between LTP and LTD of STDP need not
to be maintained precisely, but must be maintained on average.
Furthermore, we found that a basin of attraction becomes smaller
as $\delta$ increases while an initial critical overlap remains
unchanged.
\end{abstract}

%\keywords{}

\maketitle

\section{Introduction}
Recent experimental finding indicates that synaptic modification
in cortical neurons depends on the precise temporal relationship
between pre- and post-synaptic spikes
\cite[]{Bi98,Markram97,Zhang98}.
In particular, pre-synaptic spikes that precede post-synaptic
firing induce long-term potentiation (LTP) by no more than $20$
ms, while those that follow post-synaptic firing induce long-term
depression (LTD), with a rapid transition (a few ms).
The magnitude of synaptic modification decays exponentially with
the time interval between pre- and post-synaptic spikes.
This form of synaptic modification has been called
spike-timing-dependent plasticity (STDP) \cite[]{Song00} or
temporally asymmetric Hebbian learning (TAH)
\cite[]{Abbott99,Rubin01}.

The functional role of STDP has been investigated by many
authors.
They showed that STDP is a mechanism for synaptic competition
\cite[]{Abbott99,Song00,Levy01,Rossum01,Rubin01,Song01} or a
learning mechanism of sequential patterns
\cite[]{Gerstner96,Kempter99,Kistler00,Rao00,Munro00,Yoshioka02,Matsumoto02}.
Asymmetric learning window depending on spike timing like STDP has 
been studied and shown to be appropriate learning rule for
sequential patterns \cite{Herz89,Gerstner93,Abbott96}.
However, this asymmetric learning rule does not involve LTD.
Some authors showed that the balance between LTP and LTD of STDP
is significant for a stable learning
\cite[]{Song00,Munro00,Yoshioka02,Matsumoto02}.
In our previous work, we analytically showed that STDP has the
same qualitative effect as the covariance rule when the
spatio-temporal patterns are stored since the differences between
spike times that induce LTP or LTD are capable of canceling out
the effect of the firing rate information
\cite[]{Matsumoto02}.
In the brain, a situation in which the balance is maintained
precisely is inconceivable.
The data points obtained by experiments are fluctuated in the
different trials \cite{Bi98,Zhang98}.
Therefore, it is meaningful to discuss more biological plausible
situation to investigate the neuronal mechanism for sequential
learning in the brain.
Some authors \textit{numerically} investigated the impact of the
imbalance between LTP and LTD on the network properties
\cite[]{Song00,Munro00}.

The aim of this paper is to \textit{analytically} discuss the
retrieval properties of spatio-temporal patterns in an associative
memory model that incorporates the imbalance between LTP and LTD
of STDP using a method of the statistical neurodynamics
\cite[]{Matsumoto02,Okada96,Amari88}.
According to our previous work, when the balance is not precisely
maintained, it is impossible to cancel out the information of
firing rate.
Consequently, a cross-talk noise diverges.
However, if the magnitudes of LTP and LTD are equivalent on
average in a learning process, it may be possible to stably
retrieve spatio-temporal patterns.
Since the ratio of LTP and LTD is crucial, the magnitude of LTD
changes while that of LTP is fixed.
We found that the storage capacity takes a finite value even at
large $\delta$ when the fluctuation of LTD is assumed to obey a
Gaussian distribution with mean $0$ and variance $\delta^2$.
This implies that the balance between LTP and LTD of STDP need not
to be maintained precisely, but must be maintained on average.
This mechanism might work in the brain.
Furthermore, we found that a basin of attraction becomes smaller
as $\delta$ increases while an initial critical overlap remains
unchanged.

\section{Model}
The model contains $N$ binary neurons with reciprocal
connections.
Each neuron has a binary state $\{0, 1\}$.
We define discrete time steps and the following rule for
synchronous updating:
\begin{eqnarray}
u_i(t) &=& \sum_{j=1}^N J_{ij} x_j(t), \label{eq.dynamics1} \\
x_i(t+1) &=& F( u_i(t) - \theta ), \label{eq.dynamics2} \\
F(u) &=& \left\{ \begin{array}{ll}
		1. & u \geq 0 \\
		0. & u < 0, \\
		\end{array} \right.
\label{eq.dynamics3} 
\end{eqnarray}
where $x_i(t)$ is the state of the $i$-th neuron at time $t$, 
$u_i(t)$ is the internal potential of that neuron, and $\theta$ is 
a uniform threshold.
If the $i$-th neuron fires at $t$, its state is $x_i(t) = 1$;
otherwise, $x_i(t) = 0$.
$J_{ij}$ is the synaptic weight from the $j$-th neuron to the
$i$-th neuron.
Each element $\xi^{\mu}_i$ of the $\mu$-th memory pattern
$\bm{\xi}^{\mu} = (\xi^{\mu}_1, \xi^{\mu}_2, \cdots, \xi^{\mu}_N)$
is generated independently by
\begin{equation}
\mbox{Prob}[\xi^{\mu}_i = 1] = 1 - \mbox{Prob}[\xi^{\mu}_i = 0] = f. \label{eq.pattern}
\end{equation}
The expectation of $\bm{\xi}^{\mu}$ is $\mbox{E}[\xi^{\mu}_i] = f$,
and thus $f$ is considered to be the mean firing rate of
the memory pattern.
The memory pattern is sparse when $f \rightarrow 0$,
and this coding scheme is called sparse coding.

\begin{figure}[htb]
\begin{center}
\includegraphics[height = 5.4cm, width = 5.4cm]{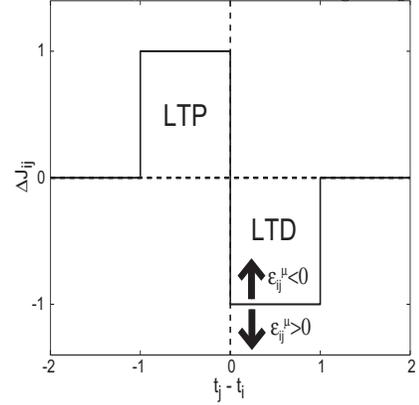}
\caption{\label{fig.asymmetric}
The time function of STDP in our model. LTP is induced
when the $j$-th neuron fires one time step before the $i$-th
one. LTD is induced when the $j$-th neuron fires one time step
after the $i$-th one. $\epsilon_{ij}^{\mu}$ follows a Gaussian
distribution with mean $\epsilon$ and variance $\delta^2$.}
\end{center}
\end{figure}

The synaptic weight $J_{ij}$ follows the form of synaptic
plasticity, which depends on the difference in spike times
between the $i$-th (post-) and $j$-th (pre-) neurons.
The time difference determines whether LTP or LTD is induced.
This type of learning rule is called spike-timing-dependent
plasticity (STDP).
The biological experimental findings show that LTP or LTD is
induced when the difference in the pre- and post-synaptic spike
times falls within about $20$ ms \cite[]{Zhang98}.
We define a single time step in equations
(\ref{eq.dynamics1}--\ref{eq.dynamics3}) as $20$ ms, and durations
within $20$ ms are ignored.
The learning rule based on STDP conforms to this equation:
\begin{equation}
J_{ij} = \frac{1}{N f (1 - f)} \sum_{\mu=1}^p \left\{ \xi_i^{\mu + 1} \xi_j^{\mu} - ( 1 + \epsilon_{ij}^{\mu-1} )\xi_i^{\mu - 1} \xi_j^{\mu}\right\}.
 \label{eq.learning}
\end{equation}
The number of memory patterns is $p= \alpha N$, 
where $\alpha$ is defined as a loading rate.
LTP is induced when the $j$-th neuron fires one time step before
the $i$-th neuron, $\xi_i^{\mu+1}=\xi_j^{\mu}=1$, while LTD is
induced when the $j$-th neuron fires one time step after the
$i$-th neuron, $\xi_i^{\mu-1}=\xi_j^{\mu}=1$.
Since the ratio of LTP and LTD is crucial, the magnitude of LTD
changes while the magnitude of LTP and the time duration are fixed.
$\epsilon_{ij}^{\mu}$ is generated independently and obeys a
Gaussian distribution with mean $\epsilon$ and variance
$\delta^2$: $\epsilon_{ij}^{\mu} \sim {\cal
N}(\epsilon,\delta^2)$.
Fig.\ref{fig.asymmetric} shows the time function of STDP in our model.
When $\epsilon_{ij}^{\mu}=0$, the balance between LTP and LTD is 
precisely maintained and then the model is equivalent to the
previous model \cite[]{Matsumoto02}.
A sequence of $p$ memory patterns is stored by STDP:
$\bm{\xi}^1 \rightarrow \bm{\xi}^2 \rightarrow \cdots
\rightarrow \bm{\xi}^p \rightarrow \bm{\xi}^1 \rightarrow
\cdots$.
In other words, $\bm{\xi}^1$ is retrieved at $t = 1$,
$\bm{\xi}^2$ is retrieved at $t = 2$, and $\bm{\xi}^{1}$ is
retrieved at $t = p+1$.
There is a critical value $\alpha_C$ of the loading rate, so that
the loading rate larger than $\alpha_C$ makes retrieval of the
pattern sequence unstable.
$\alpha_C$ is called a storage capacity.

\section{Theory}
In this section, we derive dynamical equations that describe the
retrieval properties of the network.
In this paper, we consider the thermodynamic limit: $N \rightarrow
\infty$.
The $i$-th neuronal internal potential $u_i(t)$ at time $t$ can be 
expressed as 
\begin{widetext}
\begin{eqnarray}
u_i(t) &=& \sum_{j=1}^N J_{ij} x_j(t) 
= \frac{1}{N f(1-f)} \sum_{j=1}^N \sum_{\mu=1}^p \left\{\xi_i^{\mu+1}\xi_j^{\mu} - ( 1 + \epsilon_{ij}^{\mu-1} ) \xi_i^{\mu-1} \xi_j^{\mu} \right\} x_j(t) \\
&=& \frac{1}{N f(1-f)} \sum_{j=1}^N \sum_{\mu=1}^p (\xi_i^{\mu+1}\xi_j^{\mu} - \xi_i^{\mu-1}\xi_j^{\mu} ) x_j(t) 
- \frac{1}{N f(1-f)} \sum_{j=1}^N \sum_{\mu=1}^p \epsilon_{ij}^{\mu-1} \xi_i^{\mu-1} \xi_j^{\mu} x_j(t).
\end{eqnarray}
\end{widetext}
Using the periodic boundary condition of $\xi_i^{p+1}=\xi_i^1$ and
$\xi_i^{0}=\xi_i^p$,
$\sum_{\mu=1}^p(\xi_i^{\mu+1}\xi_j^{\mu}-\xi_i^{\mu-1}\xi_j^{\mu})=\sum_{\mu=1}^p(\bar{\xi}_i^{\mu+1}\bar{\xi}_j^{\mu}-\bar{\xi}_i^{\mu-1}\bar{\xi}_j^{\mu})$ 
with $\bar{\xi}_i^{\mu}=\xi_i^{\mu}-f$. 
Using this relationship, $u_i(t)$ is given by
\begin{eqnarray}
u_i(t) &=& \frac{1}{N f(1-f)} \sum_{j=1}^N \sum_{\mu=1}^p (\bar{\xi}_i^{\mu+1} - \bar{\xi}_i^{\mu-1} ) \bar{\xi}_j^{\mu} x_j(t) \nonumber \\
& &- \frac{1}{N f(1-f)} \sum_{j=1}^N \sum_{\mu=1}^p \epsilon_{ij}^{\mu-1} \xi_i^{\mu-1} \xi_j^{\mu} x_j(t) \\
&=& \sum_{\mu=1}^{p} ( \bar{\xi}_i^{\mu+1} - \bar{\xi}_i^{\mu-1} ) m^{\mu}(t) \nonumber \\
& &- \frac{1}{N f(1-f)} \sum_{j=1}^N \sum_{\mu=1}^p \epsilon_{ij}^{\mu-1} \xi_i^{\mu-1} \xi_j^{\mu} x_j(t) \label{eq.deltau2}\\
&=& ( \bar{\xi}_i^{t+1} - \bar{\xi}_i^{t-1} ) m^{t}(t) + \sum_{\mu \neq t}^{p} ( \bar{\xi}_i^{\mu+1} - \bar{\xi}_i^{\mu-1} ) m^{\mu}(t) \nonumber \\
& &- \frac{1}{N f(1-f)} \sum_{j=1}^N \sum_{\mu=1}^p \epsilon_{ij}^{\mu-1} \xi_i^{\mu-1} \xi_j^{\mu} x_j(t)
\label{eq.deltau},
\end{eqnarray}
where $m^{\mu}(t)$ is an overlap between $\bm{\xi}^{\mu}(t)$ and
$\bm{x}(t)$ and is given by 
\begin{equation}
m^{\mu}(t)=\frac{1}{N f(1-f)} \sum_{i=1}^N \bar{\xi}_i^{\mu} x_i(t).
\end{equation}
The first term in equation (\ref{eq.deltau}) is a signal term
for the retrieval of the target pattern $\bm{\xi}^{t+1}$.
The second and third terms are a cross-talk noise term that
represents contributions from non-target patterns other than
$\bm{\xi}^{t-1}$ and that prevents $\bm{\xi}^{t+1}$ from being
retrieved.
The third term is also a compensation term originated by
a deviation from the balance between LTP and LTD of STDP.
Since this term is order of $N$ with respect to $N$, $O(N)$, when $\epsilon \neq 0$, it diverges in the thermodynamic $N$ limit: $N \rightarrow \infty$.
This means that the stored limit cycle using the present learning
rule (equation (\ref{eq.learning})) is unstable in the limit of $N 
\rightarrow \infty$ when $\epsilon \neq 0$. 
Therefore, we will discuss the $\epsilon=0$ case:
$\epsilon_{ij}^{\mu} \sim {\mathcal N}(0,\delta^2)$.

We derive the dynamical equations using the method of the statistical neurodynamics
\cite{Matsumoto02,Okada96,Amari88}.
When it is possible to store a pattern sequence, a cross-talk noise term, that is, the second and third terms in equation (\ref{eq.deltau}) is assumed to obey a Gaussian distribution with the average $0$ and time-dependent variance $\sigma^2(t)$ \cite[]{Okada96,Matsumoto02}.
We derive the recursive equations for $m^t(t)$ and $\sigma^2(t)$
to investigate whether the memory pattern $\bm{\xi}^t$ is
retrieved or not.
Since $m^t(t)$ depends on $\sigma^2(t)$, we derive $\sigma^2(t)$.
The dynamical equations are derived as
\begin{eqnarray}
m^t(t) &=& \frac{ 1 - 2f }{2} \mbox{erf}( \phi_0 ) - \frac{ 1 - f }{2} \mbox{erf}( \phi_1 ) + \frac{f}{2} \mbox{erf}( \phi_2 ),\label{eq.mt} \\
\sigma^2(t) &=& \sum_{a=0}^t {}_{2(a+1)}\mbox{C}_{(a+1)} \alpha q(t-a) \prod_{b=1}^a U^2(t-b+1) \nonumber\\
& &+ \frac{\alpha \delta^2}{(1-f)^2}q(t), \label{eq.sigmat} \\
U(t) &=& \frac{1}{\sqrt{2 \pi} \sigma(t-1)} \Bigl\{ ( 1 - 2f + 2f^2 ) e^{- \phi_0^2} \nonumber\\
& &+ f( 1 - f )( e^{- \phi_1^2} + e^{- \phi_2^2} ) \Bigr\}, \label{eq.ut}\\
q(t) &=& \frac{1}{2} \Bigl\{ 1 - ( 1 - 2f + 2f^2 ) \mbox{erf}(\phi_0) - f(1-f) ( \mbox{erf}(\phi_1) \nonumber\\
& &+ \mbox{erf}(\phi_2) ) \Bigr\}. \label{eq.qt}
\end{eqnarray}
where $\mbox{erf}(y) = \frac{2}{\sqrt{\pi}} \int_0^y \exp{(-u^2)}
du$, and $\phi_0 = \frac{\theta}{\sqrt{2}\sigma(t-1)}$,
$\phi_1 = \frac{-m^{t-1}(t-1)+\theta}{\sqrt{2} \sigma(t-1)}$,
$\phi_2 = \frac{m^{t-1}(t-1)+\theta}{\sqrt{2} \sigma(t-1)}$,
${}_{b}\mbox{C}_{a} = \frac{b!}{a! (b-a)!}$ and $a!$ is the factorial with positive integer $a$.
The detail derivation of the dynamical equations is shown in Appendix A.

\section{Results}
Fig.\ref{fig.alpha} shows the dependence of the overlap $m^t(t)$
on the loading rate $\alpha$ when the mean firing rate of the
memory pattern is $f=0.1$, and the threshold is $\theta=0.52$,
which is optimized to maximize the storage capacity.
(a) is the case at $\delta=0.0$, (b) is at $\delta=1.0$, and (c) is at $\delta=2.0$.
\begin{figure}[htb]
\begin{center}
\includegraphics{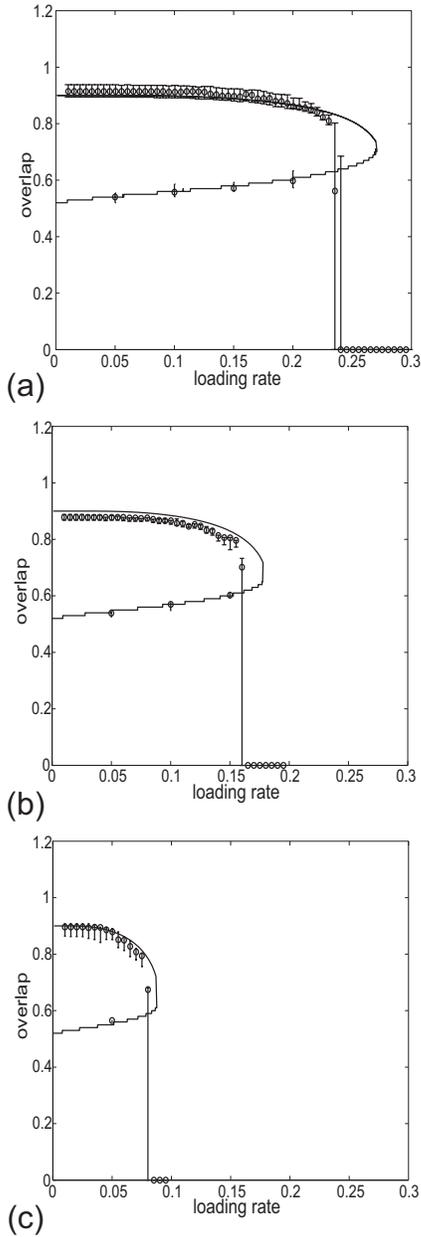}
\caption{\label{fig.alpha}
The dependence of $\alpha_C$ and the basin of attraction 
on the loading rate at $f=0.1$ and $\theta=0.52$. The lower line represents 
the initial critical overlap and the upper line does the overlap at the steady state.
The data points and the error bars show the results of computer simulation 
of $11$ trials at $N=5000$. (a):$\delta=0.0$, (b):$\delta=1.0$, (c):$\delta=2.0$.
$\alpha_C=0.27$(a), $0.178$(b), $0.087$(c).
The basin of attraction decreases as $\delta$ increases.}
\end{center}
\end{figure}
The lines show analytical results obtained by the dynamical equations (\ref{eq.mt}--\ref{eq.qt}).
The upper line denotes the steady-state values of the overlap
$m^t(t)$ in retrieval of the pattern sequence.
$m^t(t)$ is obtained by setting the initial state of the network
at the first memory pattern: $\bm{x}(1)=\bm{\xi}^1$.
Setting the initial values at $m^1(1)=1$, $\sigma^2(1)=2\alpha f + \frac{\alpha \delta^2 f}{(1-f)^2}$, $U(1)=0$ and $q(1)=f$ and using the dynamical equations (\ref{eq.mt}--\ref{eq.qt}), $m^t(t)$ is obtained.
When the overlap at the steady state is smaller than $0.5$,
the critical loading rate $\alpha$ is regarded as the storage
capacity $\alpha_C$.
The storage capacity $\alpha_C$ is $0.27$(a), $0.178$(b), and $0.087$(c).
The lower line indicates the dependence of an initial critical 
overlap $m_C$ on $\alpha$.
The stored pattern sequence is retrievable when the initial overlap
$m^1(1)$ is greater than the critical value $m_C$.
The region in which $m^1(1)$ is larger than $m_C$ represents the
basin of attraction for the retrieval of the pattern sequence.
$m_C$ is obtained by setting the initial state of the network at
$\bm{\xi}^1$ with additional noise.
We employ the following method to add noise. 
$100s\%$ of the minority components ($x_i(1)=1$) are flipped,
while the same number of majority components ($x_i(1)=0$) are also
flipped.
The initial overlap $m^1(1)$ is given as $1-\frac{2s}{1-f}$.
Then the mean firing rate of the network is kept equal to that of
the memory pattern, $f$.
The other initial values are equivalent to the upper line case.
When the overlap at the steady state is smaller than $0.5$,
the initial overlap $m^1(1)$ is regarded as the initial critical overlap $m_C$.
The data points and error bars indicate the results of computer
simulations of $11$ trials with $5000$ neurons: $N=5000$.
The results are obtained from the equations (\ref{eq.dynamics1}--\ref{eq.learning}).
The data points indicate median values and both ends of the error bars does 1/4 and 3/4 deviations.
A discrepancy between the values of $m_C$ obtained by the computer simulations and the analytical results is originated from the finite size effect of the computer simulations \cite[]{Matsumoto02}.

\begin{figure}[htb]
\begin{center}
\includegraphics[height = 5.4cm, width = 5.4cm]{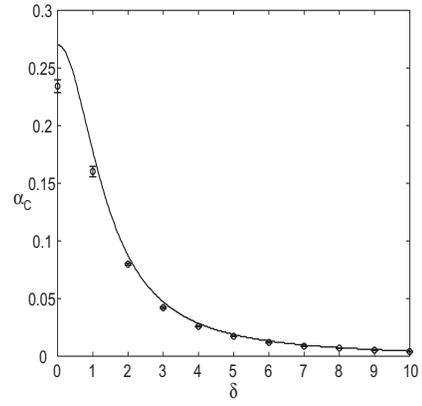}
\caption{\label{fig.deltadepend}
The dependence of the storage capacity $\alpha_C$ on $\delta$. The
solid line shows the analytical results. The data points and
error bars show the results of computer simulation of $10$ trials
at $N=5000$. Both results are obtained at $f=0.1$ and
$\theta=0.52$. As the variance $\delta^2$ of the fluctuation
increases, $\alpha_C$ decreases.}
\end{center}
\end{figure}

Fig.\ref{fig.deltadepend} shows the dependence of the storage capacity
$\alpha_C$ on the standard derivation $\delta$ of the fluctuation of LTD at $f=0.1$ and $\theta=0.52$.
The solid line shows the analytical results obtained by the same procedure to obtain $\alpha_C$ in Fig.\ref{fig.alpha}.
The data points and error bars show the results of computer simulation of $10$ trials at $N=5000$.
The means and standard deviations of $\alpha_C$ of $10$ trials are
plotted as the data points and the error bars, respectively.
As the variance $\delta^2$ increases, $\alpha_C$ decreases.
In other words, the model is robust against the imbalance
between LTP and LTD of STDP.
Thus, the balance does not need to be maintained precisely, but must
simply be maintained on average.

\begin{figure}[htb]
\begin{center}
\includegraphics[height = 5.4cm, width = 5.4cm]{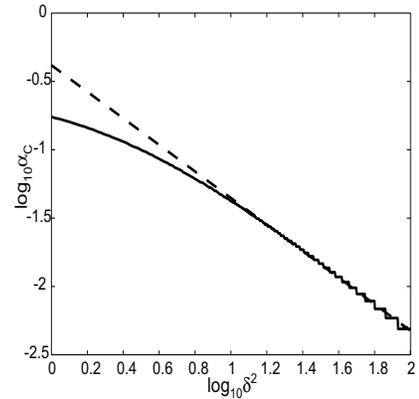}
\caption{\label{fig.limitdelta} 
An asymptotic property of $\alpha_C$ in a large limit of $\delta^2$ at
$f=0.1$ and $\theta=0.52$. The solid line shows the analytical results 
while the dashed line shows
$\log_{10}{\alpha_C}=\log_{10}{\frac{2}{\pi
\delta^2}}-1.13$. $\alpha_C$ converges to $0$ as order of $\frac{1}{\delta^2}$ with respect to $\delta$, $O(\frac{1}{\delta^2})$, in the large $\delta^2$ limit.}
\end{center}
\end{figure}

Fig.\ref{fig.limitdelta} shows an asymptotic property of $\alpha_C$ in a large limit of $\delta^2$.
The solid line shows the analytical results obtained by the same procedure to obtain $\alpha_C$ in Fig.\ref{fig.alpha} at $f=0.1$ and $\theta=0.52$ while the dashed line shows $\log_{10}{\alpha_C}=\log_{10}{\frac{2}{\pi \delta^2}}-1.13$.
This figure indicates that $\alpha_C$ converges to $0$ as order of $\frac{1}{\delta^2}$ with respect to $\delta$, $O(\frac{1}{\delta^2})$, in the large $\delta^2$ limit \cite[]{Mimura02}.

Next, we discuss the dependency of the basin of attraction on $\delta$.
Each region between the upper line and the lower line in Fig. \ref{fig.alpha}(a),(b),(c) shows the basin of attraction at $\delta=0.0,1.0,2.0$, respectively. 
The basin becomes smaller as a value of $\delta$ increases. 
However, the initial critical overlap $m_C$ is unchanged.
To introduce a threshold control scheme is known to enlarge the
basin of attraction \cite[]{Okada96,Kitano98,Matsumoto02}.

\section{Discussion}
In this paper, we investigated the impact of the imbalance between
LTP and LTD of STDP on the retrieval properties of spatio-temporal 
patterns, employing an associative memory network.
We analytically investigated the retrieval properties using
the statistical neurodynamics.
When the fluctuation of LTD is assumed to obey the Gaussian
distribution with mean $0$ and variance $\delta^2$, the storage
capacity takes a finite value even at large $\delta$.
This implies that the balance does not need to be maintained
precisely, but must be maintained on average. 
This mechanism might work in the brain.
Furthermore, the storage capacity converges to $0$ as order $O(\frac{1}{\delta^2})$ in the large $\delta^2$ limit.
Finally, we found that the basin of attraction becomes smaller as
the fluctuation of LTD increases while the initial critical
overlap remains unchanged.

We found that the storage capacity takes a finite value even at large $\delta$.
When $\delta$ is larger than $1.0$, LTD sometimes disappears in the learning process.
The spatio-temporal patterns seem not to be retrievable.
Surprisingly, even in this situation, the patterns are
retrievable.
This implies that the present model achieves strong robustness
against the imbalance between LTP and LTD.

\appendix
\section{Derivation of dynamical equations by the statistical neurodynamics}
The detail derivation of the dynamical equations (\ref{eq.mt}--\ref{eq.qt}) is given in this appendix.
At first, we give a sketch of the derivation.
The main point in this derivation is to divide an internal potential $u_i(t)$ into two parts, a signal term for a retrieval of a target pattern and a cross-talk noise term that represents contributions from non-target patterns and prevents the target pattern from being retrieved.
We evaluate the cross-talk noise term.
Specifically, the internal potential $u_i(t)$ of $i$-th neuron at time $t$ is expressed as (see equation (\ref{eq.deltau2}))
\begin{eqnarray}
u_i(t) &=& \sum_{\mu=1}^{p} ( \bar{\xi}_i^{\mu+1} - \bar{\xi}_i^{\mu-1} ) m^{\mu}(t) \nonumber \\
& &- \frac{1}{N f(1-f)} \sum_{j=1}^N \sum_{\mu=1}^p \epsilon_{ij}^{\mu-1} \xi_i^{\mu-1} \xi_j^{\mu} x_j(t) \label{eq.deltau3} \\
&=& ( \bar{\xi}_i^{t+1} - \bar{\xi}_i^{t-1} ) m^{t}(t) + \sum_{\mu \neq t}^{p} ( \bar{\xi}_i^{\mu+1} - \bar{\xi}_i^{\mu-1} ) m^{\mu}(t) \nonumber \\
& &- \frac{1}{N f(1-f)} \sum_{j=1}^N \sum_{\mu=1}^p \epsilon_{ij}^{\mu-1} \xi_i^{\mu-1} \xi_j^{\mu} x_j(t)  \label{eq.deltau4} \\
&=& ( \bar{\xi}_i^{t+1} - \bar{\xi}_i^{t-1} ) m^{t}(t) + z_i(t), \label{eq.deltau5}
\end{eqnarray}
where $\bar{\xi}_i^{\mu}=\xi_i^{\mu}-f$, $m^{\mu}(t)$ is the overlap between $\bm{\xi}^{\mu}(t)$ and $\bm{x}(t)$ and is given by 
\begin{equation}
m^{\mu}(t)=\frac{1}{N f(1-f)} \sum_{i=1}^N \bar{\xi}_i^{\mu} x_i(t), \label{eq.overlap}
\end{equation}
and 
\begin{eqnarray}
z_i(t) &=& \sum_{\mu \neq t}^{p} ( \bar{\xi}_i^{\mu+1} - \bar{\xi}_i^{\mu-1} ) m^{\mu}(t) \nonumber \\
& &- \frac{1}{N f(1-f)} \sum_{j=1}^N \sum_{\mu=1}^p \epsilon_{ij}^{\mu-1} \xi_i^{\mu-1} \xi_j^{\mu} x_j(t).
\end{eqnarray}
The first term in equation (\ref{eq.deltau5}) is the signal term and the second term is the cross-talk noise term.
Since $x_i(t)$ in equation (\ref{eq.overlap}) depends on $\xi_i^{\mu}$, the distribution of the cross-talk noise term $z_i(t)$ is unknown.
The dependence on $\xi_i^{\mu}$ is extracted from $x_i(t)$ using the Taylor expansion (see equation (\ref{eq.taylor2})).
In the thermodynamic limit, $N \rightarrow \infty$, $m^{\mu}(t)$ tends to be deterministic.
Therefore, $x_i^{\{\mu\}}(t)$, which denotes that it does not include $\xi_i^{\mu}$, is independent of $\xi_i^{\mu}$.
This enables us to assume that the cross-talk noise term $z_i(t)$ obeys a Gaussian distribution with mean $0$ and variance $\sigma^2$ \cite[]{Okada96,Matsumoto02}.
Since the distribution of the cross-talk noise term is known, the recursive equation of the overlap is obtained (see equation (\ref{eq.maverage2})).

To extract the dependence on $\xi_i^{\mu}$ from $x_i(t)$, the state of the $i$-th neuron at time $t+1$ is transformed:
\begin{eqnarray}
x_i(t+1) &=& F(u_i(t) - \theta) \nonumber \\
&=& F \Bigl( \sum_{\nu=1}^{p} ( \bar{\xi}_i^{\nu+1} - \bar{\xi}_i^{\nu-1} ) m^{\nu}(t) \nonumber \\
& &- \frac{1}{N f(1-f)} \sum_{j=1}^N \sum_{\nu=1}^p \epsilon_{ij}^{\nu-1} \xi_i^{\nu-1} \xi_j^{\nu} x_j(t) 
- \theta \Bigr).  \nonumber \\
\label{eq.xf}
\end{eqnarray}
The first term in the function $F(\cdot)$ of equation (\ref{eq.xf}) is transformed using the periodic boundary condition of $\xi^{p+1}_i=\xi^{1}_i$ and $\xi^{0}_i=\xi^{p}_i$:
\begin{eqnarray}
& & \sum_{\nu=1}^{p} ( \bar{\xi}_i^{\nu+1} - \bar{\xi}_i^{\nu-1} ) m^{\nu}(t) \nonumber \\
&=& \sum_{\nu=1}^{p} \bar{\xi}_i^{\nu+1} m^{\nu}(t) - \sum_{\nu=1}^{p} \bar{\xi}_i^{\nu-1} m^{\nu}(t) \nonumber \\
&=& \sum_{\nu'=1}^{p} \bar{\xi}_i^{\nu'} m^{\nu'-1}(t) - \sum_{\nu'=1}^{p} \bar{\xi}_i^{\nu'} m^{\nu+1}(t) \nonumber \\
&=& \sum_{\nu=1}^{p} \bar{\xi}_i^{\nu} \{ m^{\nu-1}(t) - m^{\nu+1}(t) \}. \label{eq.suffix1}
\end{eqnarray}
The second term in the function $F(\cdot)$ of equation (\ref{eq.xf}) is transformed using the periodic boundary condition:
\begin{eqnarray}
& & \frac{1}{N f(1-f)} \sum_{j=1}^N \sum_{\nu=1}^p \epsilon_{ij}^{\nu-1} \xi_i^{\nu-1} \xi_j^{\nu} x_j(t) \nonumber \\
&=& \frac{1}{N f(1-f)} \sum_{j=1}^N \sum_{\nu'=1}^p \epsilon_{ij}^{\nu'} \xi_i^{\nu'} \xi_j^{\nu'+1} x_j(t). \label{eq.suffix2}
\end{eqnarray}
To extract the dependency on $\xi_i^{\mu}$ from $x_i(t+1)$, using the equation (\ref{eq.suffix1}) and (\ref{eq.suffix2}), $x(t+1)$ is divided into two parts, the terms which include $\xi^{\mu}_i$ and the terms which do not include $\xi^{\mu}_i$:
\begin{eqnarray}
x_i(t+1)
&=& F \Bigl( \sum_{\nu=1}^{p} \bar{\xi}_i^{\nu} \left\{ m^{\nu-1}(t)-m^{\nu+1}(t)\right\} \nonumber \\
& &- \frac{1}{N f(1-f)} \sum_{j=1}^N \sum_{\nu=1}^p \epsilon_{ij}^{\nu} \xi_i^{\nu} \xi_j^{\nu+1} x_j(t) 
- \theta \Bigr) \nonumber \\
&=& F \Bigl( \bar{\xi}_i^{\mu} \left\{ m^{\mu-1}(t) - m^{\mu+1}(t) \right\} \nonumber \\
& &- \frac{1}{N f(1-f)} \sum_{j=1}^N \epsilon_{ij}^{\mu} \xi_i^{\mu} \xi_j^{\mu+1} x_j(t)\nonumber\\
& &+ \sum_{\nu \neq \mu}^{p} \bar{\xi}_i^{\nu} \left\{ m^{\nu-1}(t)-m^{\nu+1}(t)\right\} \nonumber \\
& &- \frac{1}{N f(1-f)} \sum_{j=1}^N \sum_{\nu \neq \mu}^p \epsilon_{ij}^{\nu} \xi_i^{\nu} \xi_j^{\nu+1} x_j(t) 
- \theta \Bigr). \nonumber \\
\label{eq.taylor1} 
\end{eqnarray}
At time $t$, the pattern $\bm{\xi}^t$ is designed to be retrieved.
Therefore, we can assume that $m^t(t)$ is order of $1$ with respect to $N$, $m^t(t) \sim O(1)$, and $m^{\mu}(t) (\mu \neq t)$ is order of $1/\sqrt{N}$ with respect to $N$, $m^{\mu}(t) \sim O(1/\sqrt{N})$.
Since $m^{\mu}(t) \sim O(1/\sqrt{N})$, $m^{\mu-1}(t)$ and $m^{\mu+1}(t)$ are order of $1/\sqrt{N}$ with respect to $N$.
Since $m^{\mu+1}(t) \sim O(1/\sqrt{N})$ and $\epsilon_{ij}^{\mu} \sim O(1)$, the second term in equation (\ref{eq.taylor1}) can be considered to be order of $1/\sqrt{N}$ with respect to $N$.
In the thermodynamic limit, $N \rightarrow \infty$, the first and the second terms in equation (\ref{eq.taylor1}) are small.
Using the Taylor expansion up to the first order of $x_i(t+1)$, $x_i(t+1)$ is transformed:
\begin{equation}
x_i(t+1) = x_i^{(\mu)}(t+1) + u_i^{\{\mu\}}(t) x_i'^{(\mu)}(t+1), \label{eq.taylor2}
\end{equation}
where $x_i^{(\mu)}(t+1)$ is independent of $\xi_i^{\mu}$,
$x_i'^{(\mu)}(t+1)$ is differential of $x_i^{(\mu)}(t+1)$, and 
\begin{eqnarray}
u_i^{\{\mu\}}(t) &=& \bar{\xi}_i^{\mu} \left\{ m^{\mu-1}(t) - m^{\mu+1}(t) \right\} \nonumber \\
&-& \frac{1}{N f(1-f)} \sum_{j=1}^N \epsilon_{ij}^{\mu} \xi_i^{\mu} \xi_j^{\mu+1} x_j(t), \label{eq.uextrmu} \\
x_i^{(\mu)}(t+1) &=& F(u_i(t)-u_i^{\{\mu\}}(t)-\theta), \\
x_i'^{(\mu)}(t+1) &=& F'(u_i(t)-u_i^{\{\mu\}}(t)-\theta).
\end{eqnarray}
We assume that the function $F(\cdot)$ is differentiable.
This assumption is valid since the average of $F(\cdot)$ over a Gaussian noise term will be taken in a later step (see equation (\ref{eq.maverage})).
For $\mu \neq t$, the overlap $m^{\mu}(t)$ is expressed as
\begin{widetext}
\begin{eqnarray}
m^{\mu}(t) &=&
\frac{1}{N f(1-f)} \sum_{i=1}^N (\xi_i^{\mu} - f) x_i(t)
= \frac{1}{N f(1-f)} \sum_{i=1}^N \bar{\xi}_i^{\mu} \left\{ x_i^{(\mu)}(t) + u_i^{\{\mu\}}(t-1) x_i'^{(\mu)}(t) \right\} \\
&=& \frac{1}{N f(1-f)} \sum_{i=1}^N \bar{\xi}_i^{\mu} x_i^{(\mu)}(t)
+ \frac{1}{N f(1-f)} \sum_{i=1}^N (\bar{\xi}_i^{\mu})^2 \left\{ m^{\mu-1}(t-1) - m^{\mu+1}(t-1) \right\} x_i'^{(\mu)}(t) \nonumber \\
& &- \Bigl(\frac{1}{N f(1-f)}\Bigr)^2 \sum_{i=1}^N \bar{\xi}_i^{\mu} \sum_{j=1}^N \epsilon_{ij}^{\mu} \xi_i^{\mu} \xi_j^{\mu+1} x_j(t-1) x_i'^{(\mu)}(t). \label{eq.mnott}
\end{eqnarray}
If the averages over $\xi_j^{\mu} (\mu \neq t)$ and
$\epsilon_{ji}^{\mu} (\mu \neq t)$ are taken in the right-hand
side of equation (\ref{eq.mnott}), the third term vanishes since 
$\mbox{E}[\epsilon_{ji}^{\mu}]=0$.
Since the third term including $x_j(t-1)$ depends on both $\xi_j^{\mu}$ and $\epsilon_{ji}^{\mu}$, $\xi_j^{\mu}$ and $\epsilon_{ji}^{\mu}$ are extracted from $x_j(t)$ before the averages are taken:
%\begin{widetext}
\begin{eqnarray}
x_j(t+1) &=& 
F\Bigl(\bar{\xi}_j^{\mu} \left\{ m^{\mu-1}(t) - m^{\mu+1}(t) \right\} - \frac{1}{N f(1-f)} \sum_{k \neq i}^N \epsilon_{jk}^{\mu} \xi_j^{\mu} \xi_k^{\mu+1} x_k(t) - \frac{1}{N f(1-f)} \sum_{\nu \neq \mu}^p
\epsilon_{ji}^{\nu} \xi_j^{\nu} \xi_i^{\nu+1} x_i(t) \nonumber \\
& &- \frac{1}{N f(1-f)} \epsilon_{ji}^{\mu} \xi_j^{\mu} \xi_i^{\mu+1} x_i(t) + \sum_{\nu \neq \mu}^p \bar{\xi}_j^{\nu} \left\{ m^{\nu-1}(t) - m^{\nu+1}(t) \right\}
+ \frac{1}{N f(1-f)} \sum_{k \neq i}^N \sum_{\nu \neq \mu}^p \epsilon_{jk}^{\nu} \xi_j^{\nu} \xi_k^{\nu+1} x_k(t) 
- \theta \Bigr) \nonumber \\ 
\label{eq.xmuepsilon} \\
&=& x_j^{(\mu)(\epsilon_{ji}^{\mu})}(t+1) + u_j^{\{\mu,\epsilon_{ji}^{\mu}\}}(t) x_j'^{(\mu)(\epsilon_{ji}^{\mu})}(t+1), \label{eq.xmuepsilon2}
\end{eqnarray}
%\end{widetext}
where $x_j'^{(\mu)(\epsilon_{ji}^{\mu})}(t)$, which is independent
of both $\xi_j^{\mu}(t)$ and $\epsilon_{ji}^{\mu}$, is the
differential of $x_j^{(\mu)(\epsilon_{ji}^{\mu})}(t)$ and
%\begin{widetext}
\begin{eqnarray}
u_j^{\{\mu,\epsilon_{ji}^{\mu}\}}(t) 
&=& \bar{\xi}_j^{\mu} \left\{ m^{\mu-1}(t) - m^{\mu+1}(t) \right\} 
- \frac{1}{N f(1-f)} \sum_{k \neq i}^N \epsilon_{jk}^{\mu} \xi_j^{\mu} \xi_k^{\mu+1} x_k(t) 
- \frac{1}{N f(1-f)} \sum_{\nu \neq \mu}^p \epsilon_{ji}^{\nu} \xi_j^{\nu} \xi_i^{\nu+1} x_i(t) \nonumber \\
& &- \frac{1}{N f(1-f)} \epsilon_{ji}^{\mu} \xi_j^{\mu}
\xi_i^{\mu+1} x_i(t), \label{eq.uextrepsilonmu}
\end{eqnarray}
\end{widetext}
\begin{eqnarray}
x_j^{(\mu)(\epsilon_{ji}^{\mu})}(t+1) &=&
F( u_j(t) - u_j^{\{\mu,\epsilon_{ji}^{\mu}\}}(t)-\theta), \\
x_j'^{(\mu)(\epsilon_{ji}^{\mu})}(t+1) &=&
F'( u_j(t) - u_j^{\{\mu,\epsilon_{ji}^{\mu}\}}(t)-\theta).
\end{eqnarray}
%\end{widetext}
The first, second and third term in the right-hand side of
equation (\ref{eq.uextrepsilonmu}) is order of $1/\sqrt{N}$ with respect to $N$ and the fourth term is order of $1/N$ with respect to $N$.
In the thermodynamic limit, $N \rightarrow \infty$, $u_j^{\{\mu,\epsilon_{ji}^{\mu}\}}(t)$ is small and equation (\ref{eq.xmuepsilon}) equals equation (\ref{eq.xmuepsilon2}).
Substituting equation (\ref{eq.xmuepsilon2}) into the right-hand side 
of equation (\ref{eq.mnott}) and averaging the resultant
expressions over $\xi_j^{\mu}$ and $\epsilon_{ji}^{\mu}$ and using $\mbox{E}[\epsilon_{ji}^{\mu}]=0$ yield to the following equation for $m^{\mu}(t) (\mu \neq t)$:
\begin{eqnarray}
m^{\mu}(t) &=&
\frac{1}{N f(1-f)} \sum_{i=1}^N \bar{\xi}_i^{\mu} x_i^{(\mu)}(t) \nonumber \\
& &+ \frac{1}{N f(1-f)} \sum_{i=1}^N (\bar{\xi}_i^{\mu})^2 \Big\{ m^{\mu-1}(t-1) \nonumber \\
& &- m^{\mu+1}(t-1) \Big\} x_i'^{(\mu)}(t) \\
&=& \frac{1}{N f(1-f)} \sum_{i=1}^N \bar{\xi}_i^{\mu} x_i^{(\mu)}(t) \nonumber \\
& &+ U(t) \left\{ m^{\mu-1}(t-1) - m^{\mu+1}(t-1) \right\}, \nonumber \\
\label{eq.mnott2}
\end{eqnarray}
where $U(t)=\frac{1}{N} \sum_{i=1}^N x_i'^{(\mu)}(t)$.
Substituting equation (\ref{eq.mnott2}) into equation (\ref{eq.deltau4}) yields
\begin{eqnarray}
u_i(t) &=&
( \bar{\xi}_i^{t+1} - \bar{\xi}_i^{t-1} ) m^{t}(t) \nonumber \\
& &+ \frac{1}{N f(1-f)} \sum_{j=1}^N \sum_{\mu \neq t}^p ( \bar{\xi}_i^{\mu+1} - \bar{\xi}_i^{\mu-1} ) \bar{\xi}_j^{\mu} x_j^{(\mu)}(t) \nonumber \\
& &+ \sum_{\mu \neq t}^p U(t) \Big\{ \bar{\xi}_i^{\mu+1} m^{\mu-1}(t-1) \nonumber \\
& &- 2 \bar{\xi}_i^{\mu-1} m^{\mu-1}(t-1) + \bar{\xi}_i^{\mu-1} m^{\mu+1}(t-1) \Big\} \nonumber \\
& &- \frac{1}{N f(1-f)} \sum_{j=1}^N \sum_{\mu=1}^p \epsilon_{ij}^{\mu-1} \xi_i^{\mu-1} \xi_j^{\mu} x_j(t). \label{eq.ut1}
\end{eqnarray}

Substituting equation (\ref{eq.xmuepsilon2}) into the last term of equation (\ref{eq.ut1}) yields the following expression for $u_i(t)$:
\begin{widetext}
\begin{eqnarray}
u_i(t) &=&
( \bar{\xi}_i^{t+1} - \bar{\xi}_i^{t-1} ) m^{t}(t) 
+ \frac{1}{N f(1-f)} \sum_{j=1}^N \sum_{\mu \neq t}^p ( \bar{\xi}_i^{\mu+1} - \bar{\xi}_i^{\mu-1} ) \bar{\xi}_j^{\mu} x_j^{(\mu)}(t) \nonumber \\
& &+ \sum_{\mu \neq t}^p U(t) \left\{ \bar{\xi}_i^{\mu+1} m^{\mu-1}(t-1) - 2 \bar{\xi}_i^{\mu-1} m^{\mu-1}(t-1) + \bar{\xi}_i^{\mu-1} m^{\mu+1}(t-1) \right\} \nonumber \\
& &- \frac{1}{N f(1-f)} \sum_{j=1}^N \sum_{\mu=1}^p \epsilon_{ij}^{\mu-1} \xi_i^{\mu-1} \xi_j^{\mu} \left\{ x_j^{(\mu) (\epsilon_{ji}^{\mu})}(t) + u_j^{\{\mu,\epsilon_{ji}^{\mu}\}}(t-1) x_j'^{(\mu)(\epsilon_{ji}^{\mu})}(t) \right\} \\
&=& 
( \bar{\xi}_i^{t+1} - \bar{\xi}_i^{t-1} ) m^{t}(t) 
+ \frac{1}{N f(1-f)} \sum_{j=1}^N \sum_{\mu \neq t}^p ( \bar{\xi}_i^{\mu+1} - \bar{\xi}_i^{\mu-1} ) \bar{\xi}_j^{\mu} x_j^{(\mu)}(t) \nonumber \\
& &+ \sum_{\mu \neq t}^p U(t) \left\{ \bar{\xi}_i^{\mu+1} m^{\mu-1}(t-1) - 2 \bar{\xi}_i^{\mu-1} m^{\mu-1}(t-1) + \bar{\xi}_i^{\mu-1} m^{\mu+1}(t-1) \right\} \nonumber \\
& &- \frac{1}{N f(1-f)} \sum_{j=1}^N \sum_{\mu=1}^p \epsilon_{ij}^{\mu-1} \xi_i^{\mu-1} \xi_j^{\mu} x_j^{(\mu) (\epsilon_{ji}^{\mu})}(t) \nonumber \\
& &- \frac{1}{N f(1-f)} \sum_{j=1}^N \sum_{\mu=1}^p \epsilon_{ij}^{\mu-1} \xi_i^{\mu-1} \xi_j^{\mu} \bar{\xi}_j^{\mu} \left\{ m^{\mu-1}(t-1) - m^{\mu+1}(t-1) \right\} x_j'^{(\mu) (\epsilon_{ji}^{\mu})}(t) \nonumber \\
& &+ \left( \frac{1}{N f(1-f)} \right)^2 \sum_{j=1}^N \sum_{\mu=1}^p \epsilon_{ij}^{\mu-1} \xi_i^{\mu-1} \xi_j^{\mu} \sum_{k \neq i}^N \epsilon_{jk}^{\mu} \xi_j^{\mu} \xi_k^{\mu+1} x_k(t-1) x_j'^{(\mu) (\epsilon_{ji}^{\mu})}(t) \nonumber \\
& &+ \left( \frac{1}{N f(1-f)} \right)^2 \sum_{j=1}^N \sum_{\mu=1}^p \epsilon_{ij}^{\mu-1} \xi_i^{\mu-1} \xi_j^{\mu} \sum_{\nu \neq \mu}^p \epsilon_{ji}^{\nu} \xi_j^{\nu} \xi_i^{\nu+1} x_i(t-1) x_j'^{(\mu) (\epsilon_{ji}^{\mu})}(t) \nonumber \\
& &+ \left( \frac{1}{N f(1-f)} \right)^2 \sum_{j=1}^N \sum_{\mu=1}^p \epsilon_{ij}^{\mu-1} \xi_i^{\mu-1} \xi_j^{\mu} \epsilon_{ji}^{\mu} \xi_j^{\mu} \xi_i^{\mu+1} x_i(t-1) x_j'^{(\mu) (\epsilon_{ji}^{\mu})}(t). 
\end{eqnarray}
%\end{widetext}
The fifth, sixth, seventh and eighth terms vanish since $\mbox{E}[ \epsilon_{ij}^{\mu} ] = 0$, and this yields
%\begin{widetext}
\begin{eqnarray}
u_i(t) &=&
( \bar{\xi}_i^{t+1} - \bar{\xi}_i^{t-1} ) m^{t}(t) 
+ \frac{1}{N f(1-f)} \sum_{j=1}^N \sum_{\mu \neq t}^p ( \bar{\xi}_i^{\mu+1} - \bar{\xi}_i^{\mu-1} ) \bar{\xi}_j^{\mu} x_j^{(\mu)}(t) \nonumber \\
& &+ \sum_{\mu \neq t}^p U(t) \left\{ \bar{\xi}_i^{\mu+1} m^{\mu-1}(t-1) - 2 \bar{\xi}_i^{\mu-1} m^{\mu-1}(t-1) + \bar{\xi}_i^{\mu-1} m^{\mu+1}(t-1) \right\} \nonumber \\
& &- \frac{1}{N f(1-f)} \sum_{j=1}^N \sum_{\mu=1}^p \epsilon_{ij}^{\mu-1} \xi_i^{\mu-1} \xi_j^{\mu} x_j^{(\mu) (\epsilon_{ji}^{\mu})}(t) \\
&=& ( \bar{\xi}_i^{t+1} - \bar{\xi}_i^{t-1} ) m^{t}(t) + z_i(t),
\end{eqnarray}
%\end{widetext}
where
%\begin{widetext}
\begin{eqnarray}
z_i(t) &=& 
\frac{1}{N f(1-f)} \sum_{j=1}^N \sum_{\mu \neq t}^p ( \bar{\xi}_i^{\mu+1} - \bar{\xi}_i^{\mu-1} ) \bar{\xi}_j^{\mu} x_j^{(\mu)}(t)
+ \sum_{\mu \neq t}^p U(t) \Bigl\{ \bar{\xi}_i^{\mu+1} m^{\mu-1}(t-1) - 2 \bar{\xi}_i^{\mu-1} m^{\mu-1}(t-1) \nonumber\\
& &+ \bar{\xi}_i^{\mu-1} m^{\mu+1}(t-1) \Bigr\} 
- \frac{1}{N f(1-f)} \sum_{j=1}^N \sum_{\mu=1}^p \epsilon_{ij}^{\mu-1} \xi_i^{\mu-1} \xi_j^{\mu} x_j^{(\mu) (\epsilon_{ji}^{\mu})}(t).
\end{eqnarray}
%\end{widetext}
$z_i(t)$ is the cross-talk noise term.
We assume that the cross-talk noise obeys a Gaussian distribution
with mean $0$ and time-dependent variance $\sigma^2(t)$:
$\mbox{E}[ z_i(t) ] =0, \mbox{E} [ (z_i(t))^2 ] = \sigma^2(t)$ \cite[]{Okada96,Matsumoto02}.
The first and second term of $z_i(t)$ are the same
cross-talk noise term as that of our previous work \cite{Matsumoto02}.
The square of $z_i(t)$ is given by
%\begin{widetext}
\begin{eqnarray}
\left\{z_i(t)\right\}^2 &=& 
\left( \frac{1}{N f(1-f)} \right)^2 \sum_{j=1}^N \sum_{\mu \neq t}^{p} ( \bar{\xi}_i^{\mu+1} - \bar{\xi}_i^{\mu-1} )^2 ( \bar{\xi}_j^{\mu} )^2 \left\{ x_j^{(\mu)}(t) \right\}^2 \nonumber \\
& &+ \sum_{\nu \neq t}^{p} U(t)^2 \Bigl\{ \bar{\xi}_i^{\nu+1} m^{\nu-1}(t-1) - 2 \bar{\xi}_i^{\nu-1} m^{\nu-1}(t-1) + \bar{\xi}_i^{\nu-1} m^{\nu+1}(t-1) \Bigr\}^2 \nonumber \\
& &+ \left(\frac{1}{N f(1-f)} \right)^2 \sum_{j=1}^N \sum_{\mu=1}^p (\epsilon_{ij}^{\mu-1})^2 (\xi_i^{\mu-1})^2 (\xi_j^{\mu})^2 \left\{ x_j^{(\mu)(\epsilon_{ji}^{\mu})}(t) \right\}^2 \nonumber \\
& &+ \frac{2}{N f(1-f)} \sum_{j=1}^N \sum_{\mu \neq t}^p ( \bar{\xi}_i^{\mu+1} - \bar{\xi}_i^{\mu-1} ) \bar{\xi}_j^{\mu} x_j^{(\mu)}(t) \nonumber \\
& &\times \sum_{\mu' \neq t}^p U(t) \Bigl\{ \bar{\xi}_i^{\mu'+1} m^{\mu'-1}(t-1) - 2 \bar{\xi}_i^{\mu'-1} m^{\mu'-1}(t-1) + \bar{\xi}_i^{\mu'-1} m^{\mu'+1}(t-1) \Bigr\} \nonumber \\ 
& &+2 \sum_{\mu \neq t}^p U(t) \Bigl\{ \bar{\xi}_i^{\mu+1} m^{\mu-1}(t-1) - 2 \bar{\xi}_i^{\mu-1} m^{\mu-1}(t-1) + \bar{\xi}_i^{\mu-1} m^{\mu+1}(t-1) \Bigr\} \nonumber \\ 
& &\times \frac{1}{N f(1-f)} \sum_{j=1}^N \sum_{\mu=1}^p \epsilon_{ij}^{\mu-1} \xi_i^{\mu-1} \xi_j^{\mu} x_j^{(\mu) (\epsilon_{ji}^{\mu})}(t) \nonumber \\
& &+2 \left( \frac{1}{N f(1-f)} \right)^2 \sum_{j=1}^N \sum_{\mu'=1}^p \epsilon_{ij}^{\mu'-1} \xi_i^{\mu'-1} \xi_j^{\mu'} x_j^{(\mu') (\epsilon_{ji}^{\mu'})}(t)
\sum_{k=1}^N \sum_{\mu' \neq t}^p ( \bar{\xi}_i^{\mu'+1} - \bar{\xi}_i^{\mu'-1} ) \bar{\xi}_k^{\mu'} x_k^{(\mu')}(t) \label{eq.z2} \\
&=& \left( \frac{1}{N f(1-f)} \right)^2 \sum_{j=1}^N \sum_{\mu \neq t}^{p} ( \bar{\xi}_i^{\mu+1} - \bar{\xi}_i^{\mu-1} )^2 ( \bar{\xi}_j^{\mu} )^2 \left\{ x_j^{(\mu)}(t) \right\}^2 \nonumber \\
& &+ \sum_{\nu \neq t}^{p} U(t)^2 \Bigl\{ \bar{\xi}_i^{\nu+1} m^{\nu-1}(t-1) - 2 \bar{\xi}_i^{\nu-1} m^{\nu-1}(t-1) + \bar{\xi}_i^{\nu-1} m^{\nu+1}(t-1) \Bigr\}^2 \nonumber \\
& &+ \left( \frac{1}{N f(1-f)} \right)^2 \sum_{j=1}^N \sum_{\mu=1}^p (\epsilon_{ij}^{\mu-1})^2 (\xi_i^{\mu-1})^2 (\xi_j^{\mu})^2 \left\{ x_j^{(\mu)(\epsilon_{ji}^{\mu})}(t) \right\}^2 \\
&=& \sum_{a=0}^t {}_{2(a+1)}\mbox{C}_{(a+1)} \alpha q(t-a) \prod_{b=1}^a U^2(t-b+1) 
+ \frac{\alpha \delta^2}{(1-f)^2}q(t),
\end{eqnarray}
\end{widetext}
where $p = \alpha N$, $q(t) = \frac{1}{N} \sum_{i=1}^N
\left\{ x_i^{(\mu)}(t) \right\}^2$, ${}_{b}\mbox{C}_{a} =
\frac{b!}{a! (b-a)!}$, $a!$ is the factorial with positive integer
$a$ and $\mbox{E}[ (\epsilon_{ij}^{\mu})^2 ]=\delta^2$.
Since $\mbox{E}[\epsilon_{ij}^{\mu}]=0$, the fourth, fifth and
sixth terms in equation (\ref{eq.z2}) vanish.
We applied the relationship $\sum_{a=0}^b({}_{b}\mbox{C}_{a})^2 =
{}_{2b}\mbox{C}_{b}$ in this derivation.
Since $\mbox{E}[ z_i(t) ] = 0$, the variance
$\sigma^2(t)$ is equal to $\mbox{E}[\{z_i(t)\}^2]$.
We then get the recursive equation for $\sigma^2(t)$:
\begin{eqnarray}
\sigma^2(t) &=& \sum_{a=0}^t {}_{2(a+1)}\mbox{C}_{(a+1)} \alpha q(t-a) \prod_{b=1}^a U^2(t-b+1) \nonumber\\
& &+ \frac{\alpha \delta^2}{(1-f)^2}q(t).
\end{eqnarray}
The overlap between the state $\bm{x}(t)$ and the retrieval pattern $\bm{\xi}^{t}$ is given by
%\begin{widetext}
\begin{eqnarray}
m^t(t) 
&=& \frac{1}{Nf(1-f)} \sum_{i=1}^N ( \xi_i^t - f ) x_i(t) \nonumber \\
&=& \frac{1}{Nf(1-f)} \sum_{i=1}^N ( \xi_i^t - f ) \nonumber \\
& & \times F\Bigl((\xi_i^t - \xi_i^{t-2}) m^{t-1}(t-1) + z_i(t-1) - \theta \Bigr). \nonumber 
\\
\end{eqnarray}
Since $u_i(t)$ is independent and identical distribution (i.i.d.), by the law of large numbers, the average over $i$ can be replaced by an average over the memory patterns $\xi^{\mu}$ and the Gaussian noise term $z \sim \mathcal{N}(0, \sigma^2)$.
Then, the recursive equation for the overlap $m^t(t)$ is transformed:
\begin{eqnarray}
m^t(t) &=& 
\frac{1}{f(1-f)} \frac{1}{\sqrt{2 \pi} \sigma} \int_{-\infty}^{\infty} dz e^{-\frac{z^2}{2 \sigma^2}} \Bigl\langle \Bigl\langle ( \xi^t - f ) \nonumber \\
& & \times F\Bigl((\xi^t - \xi^{t-2}) m^{t-1}(t-1) + z
- \theta \Bigr) \Bigr\rangle \Bigr\rangle \nonumber\\
&=& \frac{1}{f(1-f)} \frac{1}{\sqrt{2 \pi}} \int_{-\infty}^{\infty} d\tilde{z} e^{-\frac{\tilde{z}^2}{2}} \Bigl\langle
\Bigl\langle ( \xi^t - f ) \nonumber\\
& &\times F\Bigl((\xi^t - \xi^{t-2}) m^{t-1}(t-1) + \sigma(t-1) \tilde{z}
- \theta \Bigr) \Bigr\rangle \Bigr\rangle \nonumber\\
\label{eq.maverage} \\
&=& \frac{ 1 - 2f }{2} \mbox{erf}( \phi_0 ) - \frac{ 1 - f }{2} \mbox{erf}( \phi_1 ) + \frac{f}{2} \mbox{erf}( \phi_2 ), \label{eq.maverage2}
\end{eqnarray}
%\end{widetext}
where $\tilde{z}=z/\sigma$, $\mbox{erf}(y) = \frac{2}{\sqrt{\pi}} \int_0^y \exp{(-u^2)}
du$, and $\phi_0 = \frac{\theta}{\sqrt{2}\sigma(t-1)}$,
$\phi_1 = \frac{-m^{t-1}(t-1)+\theta}{\sqrt{2} \sigma(t-1)}$,
$\phi_2 = \frac{m^{t-1}(t-1)+\theta}{\sqrt{2} \sigma(t-1)}$, and $\langle \langle \cdot \rangle \rangle$ denotes an average over the memory pattern $\xi^{\mu}$.
Since $x_i'(t)-x_i'^{(\mu)}(t) \sim O(\frac{1}{\sqrt{N}})$ and the thermodynamic limit, $N \rightarrow \infty$, is considered, $x_i'^{(\mu)}(t)=x_i'(t)$.
Using this relationship, we derive $U(t)$:
%\begin{widetext}
\begin{eqnarray}
U(t) &=& \frac{1}{N} \sum_{i=1}^N x_i'^{(\mu)}(t) = \frac{1}{N} \sum_{i=1}^N x_i'(t) \\
&=& \frac{1}{f(1-f)} \frac{1}{\sqrt{2 \pi}} \int_{-\infty}^{\infty} dz e^{-\frac{z^2}{2}} \nonumber \\
& &\times \Bigl\langle \Bigl\langle F'\Bigl( ( \xi^t - \xi^{t-2} ) m^{t-1}(t-1) \nonumber\\
& &+ \sigma(t-1) z - \theta \Bigr) \Bigr\rangle \Bigr\rangle \\
&=& \frac{1}{f(1-f)} \frac{1}{\sqrt{2 \pi}} \int_{-\infty}^{\infty} dz e^{-\frac{z^2}{2}} z \nonumber \\
& &\times \Bigl\langle \Bigl\langle F\Bigl( ( \xi^t - \xi^{t-2} ) m^{t-1}(t-1) \nonumber\\
& &+ \sigma(t-1) z - \theta \Bigr) \Bigr\rangle \Bigr\rangle \\
&=& \frac{1}{\sqrt{2 \pi} \sigma(t-1)} \Bigl\{ ( 1 - 2f + 2f^2 ) e^{- \phi_0^2} \nonumber\\
& &+ f( 1 - f )( e^{- \phi_1^2} + e^{- \phi_2^2} ) \Bigr\}.
\end{eqnarray}
%\end{widetext}
Since $x_i(t)-x_i^{(\mu)}(t) \sim O(\frac{1}{\sqrt{N}})$ and $N \rightarrow \infty$, $x_i^{(\mu)}(t)=x_i(t)$.
Using this relationship, we derive $q(t)$:
%\begin{widetext}
\begin{eqnarray}
q(t) &=& \frac{1}{N} \sum_{i=1}^N \{x_i^{(\mu)}(t)\}^2 = \frac{1}{N} \sum_{i=1}^N \{x_i(t)\}^2 \\
&=& \frac{1}{f(1-f)} \frac{1}{\sqrt{2 \pi}} \int_{-\infty}^{\infty} dz e^{-\frac{z^2}{2}} \nonumber \\
& &\times \Bigl\langle \Bigl\langle F^2\Bigl( ( \xi^t - \xi^{t-2} ) m^{t-1}(t-1) \nonumber\\
& &+ \sigma(t-1)z - \theta \Bigr) \Bigr\rangle \Bigr\rangle \\
&=& \frac{1}{2} \Bigl\{ 1 - ( 1 - 2f + 2f^2 ) \mbox{erf}(\phi_0) - f(1-f) ( \mbox{erf}(\phi_1) \nonumber\\
& &+ \mbox{erf}(\phi_2) ) \Bigr\}.
\end{eqnarray}
%\end{widetext}


\begin{thebibliography}{23}
\expandafter\ifx\csname natexlab\endcsname\relax\def\natexlab#1{#1}\fi
\expandafter\ifx\csname bibnamefont\endcsname\relax
  \def\bibnamefont#1{#1}\fi
\expandafter\ifx\csname bibfnamefont\endcsname\relax
  \def\bibfnamefont#1{#1}\fi
\expandafter\ifx\csname citenamefont\endcsname\relax
  \def\citenamefont#1{#1}\fi
\expandafter\ifx\csname url\endcsname\relax
  \def\url#1{\texttt{#1}}\fi
\expandafter\ifx\csname urlprefix\endcsname\relax\def\urlprefix{URL }\fi
\providecommand{\bibinfo}[2]{#2}
\providecommand{\eprint}[2][]{\url{#2}}

\bibitem[{\citenamefont{Bi and Poo}(1998)}]{Bi98}
\bibinfo{author}{\bibfnamefont{G.~Q.} \bibnamefont{Bi}} \bibnamefont{and}
  \bibinfo{author}{\bibfnamefont{M.~M.} \bibnamefont{Poo}},
  \bibinfo{journal}{Journal of Neuroscience} \textbf{\bibinfo{volume}{18}},
  \bibinfo{pages}{10464} (\bibinfo{year}{1998}).

\bibitem[{\citenamefont{Markram et~al.}(1997)\citenamefont{Markram, L\"{u}bke,
  Frotscher, and Sakmann}}]{Markram97}
\bibinfo{author}{\bibfnamefont{H.}~\bibnamefont{Markram}},
  \bibinfo{author}{\bibfnamefont{J.}~\bibnamefont{L\"{u}bke}},
  \bibinfo{author}{\bibfnamefont{M.}~\bibnamefont{Frotscher}},
  \bibnamefont{and} \bibinfo{author}{\bibfnamefont{B.}~\bibnamefont{Sakmann}},
  \bibinfo{journal}{Science} \textbf{\bibinfo{volume}{275}},
  \bibinfo{pages}{213} (\bibinfo{year}{1997}).

\bibitem[{\citenamefont{Zhang et~al.}(1998)\citenamefont{Zhang, Tao, Holt,
  Harris, and Poo}}]{Zhang98}
\bibinfo{author}{\bibfnamefont{L.~I.} \bibnamefont{Zhang}},
  \bibinfo{author}{\bibfnamefont{H.~W.} \bibnamefont{Tao}},
  \bibinfo{author}{\bibfnamefont{C.~E.} \bibnamefont{Holt}},
  \bibinfo{author}{\bibfnamefont{W.~A.} \bibnamefont{Harris}},
  \bibnamefont{and} \bibinfo{author}{\bibfnamefont{M.~M.} \bibnamefont{Poo}},
  \bibinfo{journal}{Nature} \textbf{\bibinfo{volume}{395}}, \bibinfo{pages}{37}
  (\bibinfo{year}{1998}).

\bibitem[{\citenamefont{Song et~al.}(2000)\citenamefont{Song, Miller, and
  Abbott}}]{Song00}
\bibinfo{author}{\bibfnamefont{S.}~\bibnamefont{Song}},
  \bibinfo{author}{\bibfnamefont{K.~D.} \bibnamefont{Miller}},
  \bibnamefont{and} \bibinfo{author}{\bibfnamefont{L.~F.}
  \bibnamefont{Abbott}}, \bibinfo{journal}{Nature Neuroscience}
  \textbf{\bibinfo{volume}{3}}, \bibinfo{pages}{919} (\bibinfo{year}{2000}).

\bibitem[{\citenamefont{Abbott and Song}(1999)}]{Abbott99}
\bibinfo{author}{\bibfnamefont{L.~F.} \bibnamefont{Abbott}} \bibnamefont{and}
  \bibinfo{author}{\bibfnamefont{S.}~\bibnamefont{Song}}, in
  \emph{\bibinfo{booktitle}{Advances in Neural Information Processing
  Systems}}, edited by \bibinfo{editor}{\bibfnamefont{M.~S.}
  \bibnamefont{Kearns}},
  \bibinfo{editor}{\bibfnamefont{S.}~\bibnamefont{Solla}}, \bibnamefont{and}
  \bibinfo{editor}{\bibfnamefont{D.}~\bibnamefont{Cohn}}
  (\bibinfo{publisher}{Cambridge, MA:MIT Press}, \bibinfo{year}{1999}),
  vol.~\bibinfo{volume}{11}, pp. \bibinfo{pages}{69--75}.

\bibitem[{\citenamefont{Rubin et~al.}(2001)\citenamefont{Rubin, Lee, and
  Sompolinsky}}]{Rubin01}
\bibinfo{author}{\bibfnamefont{J.}~\bibnamefont{Rubin}},
  \bibinfo{author}{\bibfnamefont{D.~D.} \bibnamefont{Lee}}, \bibnamefont{and}
  \bibinfo{author}{\bibfnamefont{H.}~\bibnamefont{Sompolinsky}},
  \bibinfo{journal}{Physical Review Letters} \textbf{\bibinfo{volume}{86}},
  \bibinfo{pages}{364} (\bibinfo{year}{2001}).

\bibitem[{\citenamefont{Levy et~al.}(2001)\citenamefont{Levy, Horn, Meilijson,
  and Ruppin}}]{Levy01}
\bibinfo{author}{\bibfnamefont{N.}~\bibnamefont{Levy}},
  \bibinfo{author}{\bibfnamefont{D.}~\bibnamefont{Horn}},
  \bibinfo{author}{\bibfnamefont{I.}~\bibnamefont{Meilijson}},
  \bibnamefont{and} \bibinfo{author}{\bibfnamefont{E.}~\bibnamefont{Ruppin}},
  \bibinfo{journal}{Nerural Networks} \textbf{\bibinfo{volume}{14}},
  \bibinfo{pages}{815} (\bibinfo{year}{2001}).

\bibitem[{\citenamefont{van Rossum et~al.}(2001)\citenamefont{van Rossum, Bi,
  and Turrigiano}}]{Rossum01}
\bibinfo{author}{\bibfnamefont{M.~C.~W.} \bibnamefont{van Rossum}},
  \bibinfo{author}{\bibfnamefont{G.~Q.} \bibnamefont{Bi}}, \bibnamefont{and}
  \bibinfo{author}{\bibfnamefont{G.~G.} \bibnamefont{Turrigiano}},
  \bibinfo{journal}{Journal of Neuroscience} \textbf{\bibinfo{volume}{20}},
  \bibinfo{pages}{8812} (\bibinfo{year}{2001}).

\bibitem[{\citenamefont{Song and Abbott}(2001)}]{Song01}
\bibinfo{author}{\bibfnamefont{S.}~\bibnamefont{Song}} \bibnamefont{and}
  \bibinfo{author}{\bibfnamefont{L.~F.} \bibnamefont{Abbott}},
  \bibinfo{journal}{Neuron} \textbf{\bibinfo{volume}{32}}, \bibinfo{pages}{339}
  (\bibinfo{year}{2001}).

\bibitem[{\citenamefont{Gerstner et~al.}(1996)\citenamefont{Gerstner, Kempter,
  van Hemmen, and Wagner}}]{Gerstner96}
\bibinfo{author}{\bibfnamefont{W.}~\bibnamefont{Gerstner}},
  \bibinfo{author}{\bibfnamefont{R.}~\bibnamefont{Kempter}},
  \bibinfo{author}{\bibfnamefont{J.~L.} \bibnamefont{van Hemmen}},
  \bibnamefont{and} \bibinfo{author}{\bibfnamefont{H.}~\bibnamefont{Wagner}},
  \bibinfo{journal}{Nature} \textbf{\bibinfo{volume}{383}}, \bibinfo{pages}{76}
  (\bibinfo{year}{1996}).

\bibitem[{\citenamefont{Kempter et~al.}(1999)\citenamefont{Kempter, Gerstner,
  and van Hemmen}}]{Kempter99}
\bibinfo{author}{\bibfnamefont{R.}~\bibnamefont{Kempter}},
  \bibinfo{author}{\bibfnamefont{W.}~\bibnamefont{Gerstner}}, \bibnamefont{and}
  \bibinfo{author}{\bibfnamefont{J.~L.} \bibnamefont{van Hemmen}},
  \bibinfo{journal}{Physical Review E} \textbf{\bibinfo{volume}{59}},
  \bibinfo{pages}{4498} (\bibinfo{year}{1999}).

\bibitem[{\citenamefont{Kistler and van Hemmen}(2000)}]{Kistler00}
\bibinfo{author}{\bibfnamefont{W.~M.} \bibnamefont{Kistler}} \bibnamefont{and}
  \bibinfo{author}{\bibfnamefont{J.~L.} \bibnamefont{van Hemmen}},
  \bibinfo{journal}{Neural Computation} \textbf{\bibinfo{volume}{12}},
  \bibinfo{pages}{385} (\bibinfo{year}{2000}).

\bibitem[{\citenamefont{Rao and Sejnowski}(2000)}]{Rao00}
\bibinfo{author}{\bibfnamefont{R.~P.~N.} \bibnamefont{Rao}} \bibnamefont{and}
  \bibinfo{author}{\bibfnamefont{T.~J.} \bibnamefont{Sejnowski}}, in
  \emph{\bibinfo{booktitle}{Advances in Neural Information Processing
  Systems}}, edited by \bibinfo{editor}{\bibfnamefont{S.~A.}
  \bibnamefont{Solla}}, \bibinfo{editor}{\bibfnamefont{T.~K.}
  \bibnamefont{Leen}}, \bibnamefont{and} \bibinfo{editor}{\bibfnamefont{K.~R.}
  \bibnamefont{M\"{u}ller}} (\bibinfo{publisher}{Cambridge, MA:MIT Press},
  \bibinfo{year}{2000}), vol.~\bibinfo{volume}{12}, pp.
  \bibinfo{pages}{164--170}.

\bibitem[{\citenamefont{Munro and Hernandez}(2000)}]{Munro00}
\bibinfo{author}{\bibfnamefont{P.}~\bibnamefont{Munro}} \bibnamefont{and}
  \bibinfo{author}{\bibfnamefont{G.}~\bibnamefont{Hernandez}}, in
  \emph{\bibinfo{booktitle}{Advances in Neural Information Processing
  Systems}}, edited by \bibinfo{editor}{\bibfnamefont{S.~A.}
  \bibnamefont{Solla}}, \bibinfo{editor}{\bibfnamefont{T.~K.}
  \bibnamefont{Leen}}, \bibnamefont{and} \bibinfo{editor}{\bibfnamefont{K.~R.}
  \bibnamefont{M\"{u}ller}} (\bibinfo{publisher}{Cambridge, MA:MIT Press},
  \bibinfo{year}{2000}), vol.~\bibinfo{volume}{12}, pp.
  \bibinfo{pages}{150--156}.

\bibitem[{\citenamefont{Yoshioka}(2002)}]{Yoshioka02}
\bibinfo{author}{\bibfnamefont{M.}~\bibnamefont{Yoshioka}},
  \bibinfo{journal}{Physical Review E} \textbf{\bibinfo{volume}{65}},
  \bibinfo{pages}{011903} (\bibinfo{year}{2002}).

\bibitem[{\citenamefont{Matsumoto and Okada}(2002)}]{Matsumoto02}
\bibinfo{author}{\bibfnamefont{N.}~\bibnamefont{Matsumoto}} \bibnamefont{and}
  \bibinfo{author}{\bibfnamefont{M.}~\bibnamefont{Okada}},
  \bibinfo{journal}{Neural Computation} \textbf{\bibinfo{volume}{14}},
  \bibinfo{pages}{2883} (\bibinfo{year}{2002}), \bibinfo{note}{in
  \textit{Advances in Neural Information Processing Systems}, edited by T. G.
  Dietterich, S. Becker and Z. Ghahramani (Cambridge, MA:MIT Press, 2002), vol.
  14, 245--252}.

\bibitem[{\citenamefont{Herz et~al.}(1989)\citenamefont{Herz, Sulzer, Kuhn, and
  van Hemmen}}]{Herz89}
\bibinfo{author}{\bibfnamefont{A.}~\bibnamefont{Herz}},
  \bibinfo{author}{\bibfnamefont{B.}~\bibnamefont{Sulzer}},
  \bibinfo{author}{\bibfnamefont{R.}~\bibnamefont{Kuhn}}, \bibnamefont{and}
  \bibinfo{author}{\bibfnamefont{J.~L.} \bibnamefont{van Hemmen}},
  \bibinfo{journal}{Biological Cybernetics} \textbf{\bibinfo{volume}{60}},
  \bibinfo{pages}{457} (\bibinfo{year}{1989}).

\bibitem[{\citenamefont{Gerstner et~al.}(1993)\citenamefont{Gerstner, Ritz, and
  van Hemmen}}]{Gerstner93}
\bibinfo{author}{\bibfnamefont{W.}~\bibnamefont{Gerstner}},
  \bibinfo{author}{\bibfnamefont{R.}~\bibnamefont{Ritz}}, \bibnamefont{and}
  \bibinfo{author}{\bibfnamefont{J.~L.} \bibnamefont{van Hemmen}},
  \bibinfo{journal}{Biological Cybernetic} \textbf{\bibinfo{volume}{69}},
  \bibinfo{pages}{503} (\bibinfo{year}{1993}).

\bibitem[{\citenamefont{Abbott and Blum}(1996)}]{Abbott96}
\bibinfo{author}{\bibfnamefont{L.~F.} \bibnamefont{Abbott}} \bibnamefont{and}
  \bibinfo{author}{\bibfnamefont{K.~I.} \bibnamefont{Blum}},
  \bibinfo{journal}{Cerebral Cortex} \textbf{\bibinfo{volume}{6}},
  \bibinfo{pages}{406} (\bibinfo{year}{1996}).

\bibitem[{\citenamefont{Okada}(1996)}]{Okada96}
\bibinfo{author}{\bibfnamefont{M.}~\bibnamefont{Okada}},
  \bibinfo{journal}{Neural Networks} \textbf{\bibinfo{volume}{9}},
  \bibinfo{pages}{1429} (\bibinfo{year}{1996}).

\bibitem[{\citenamefont{Amari and Maginu}(1988)}]{Amari88}
\bibinfo{author}{\bibfnamefont{S.}~\bibnamefont{Amari}} \bibnamefont{and}
  \bibinfo{author}{\bibfnamefont{K.}~\bibnamefont{Maginu}},
  \bibinfo{journal}{Neural Networks} \textbf{\bibinfo{volume}{1}},
  \bibinfo{pages}{63} (\bibinfo{year}{1988}).

\bibitem[{\citenamefont{Mimura et~al.}()\citenamefont{Mimura, Kimoto, and
  Okada}}]{Mimura02}
\bibinfo{author}{\bibfnamefont{K.}~\bibnamefont{Mimura}},
  \bibinfo{author}{\bibfnamefont{T.}~\bibnamefont{Kimoto}}, \bibnamefont{and}
  \bibinfo{author}{\bibfnamefont{M.}~\bibnamefont{Okada}},
  \bibinfo{note}{cond-mat/0207545}.

\bibitem[{\citenamefont{Kitano and Aoyagi}(1998)}]{Kitano98}
\bibinfo{author}{\bibfnamefont{K.}~\bibnamefont{Kitano}} \bibnamefont{and}
  \bibinfo{author}{\bibfnamefont{T.}~\bibnamefont{Aoyagi}},
  \bibinfo{journal}{Journal of Physics A: Mathematical and General}
  \textbf{\bibinfo{volume}{31}}, \bibinfo{pages}{L613} (\bibinfo{year}{1998}).

\end{thebibliography}
\end{document}